\documentclass[reprint,prb,showpacs,showkeys]{revtex4-1}

\usepackage[T1]{fontenc}
\pdfoutput=1
\usepackage{amsmath}
\usepackage{graphicx}

\usepackage[unicode=true,pdfusetitle,
 bookmarks=true,bookmarksnumbered=false,bookmarksopen=false,
 breaklinks=false,pdfborder={0 0 0},pdfborderstyle={},backref=false,colorlinks=true]
 {hyperref}

\begin{document}

\title{Effect of van Hove singularity on the isotope effect and critical temperature of H$_{3}$S  hydride superconductor as a function of pressure}

\author{S. Villa-Cort{\'e}s and O. De la Pe{\~n}a-Seaman}

\email{svilla@ifuap.buap.mx}

\affiliation{Instituto de F{\'i}sica, Benem{\'e}rita Universidad Aut{\'o}noma de Puebla, Apartado Postal J-48, 72570, Puebla, Puebla, M{\'e}xico }

\date{\today}

\begin{abstract}
We have performed density functional calculations in conjunction with the linearized Migdal-Eliashberg equations and the functional derivative approach, which takes into account the energy-range dependence of the density of states at the Fermi level ($N(\varepsilon)$ variable on the scale of phonon energy), to determine the evolution of the critical temperature ($T_{c}$) and the isotope effect coefficient ($\alpha$) of H$_{3}$S as a function of pressure on its $lm\bar{3}m$ crystal-structure range (from 162 to 250 GPa). Such approach, in comparison with $N(\varepsilon)$=$N(0)$=const., improves the agreement of $T_{c}$ with available experiments on the whole range of studied pressure. Considering for $\alpha$ two main contributions: one positive coming from the electron-phonon (el-ph) and the other negative from the electron-electron interaction (el-el), we obtained a monotonic decrement as a function of pressure, independent of applied scheme ($N(\varepsilon)$ or $N(0)$). However, when $N(\varepsilon)$ is taken into account, an important renormalization occurs on both contributions, el-ph and el-el, improving the agreement with experimental data, specially for the high-pressure regime. The observed evolution of $T_{c}$ and $\alpha$ as a function of pressure indicates, thus, the crucial role of the energy-dependence on $N(\varepsilon)$ for the proper analysis and description of the superconducting state on high-$T_{c}$ metal hydrides as H$_{3}$S, by considering the role of its van Hove singularities. 
\end{abstract}




\maketitle

\section{Introduction}

Since Kamerlingh Onnes discovered the superconductivity at 4.2 K in Mercury in 1911 \cite{Onnes_1}, it has been a challenge for theoretical physics to predict as well as understand the mechanism responsible for Cooper pair formation in superconducting materials. In the last decade, several theoretical predictions have been made on the crystal structure of stoichiometric and hydrogen-rich compounds at high pressures for which their electronic, dynamic and coupling properties had been calculated \cite{cor6,Duan2019,PhysRevB.96.100502,BI2019}. As a result from those predictions, several metal hydrides have been proposed as conventional-superconductor candidates with a superconducting critical temperature ($T_{c}$) near to room temperature \cite{10.1038/srep06968,Liu6990,Wang6463}. The astonishing experimental observation of conventional superconductivity at $T_{c}=203$ K in H$_{3}$S under a hydrostatic pressure of 155 GPa \cite{203} and more recently at similar pressures in LaH$_{10}$ with $T_{c}$ in the range of 250-260 K \cite{La50,PhysRevLett.122.027001}, raises the possibility that more metal-hydrides could be discovered as superconductors, and some of them even with higher transition temperatures. In order to make predictions in new materials that could be synthesized in a laboratory, it is desirable to establish general characteristics that allow the understanding of the underlying mechanisms of high-$T_{c}$ superconductivity in hydrogen-rich materials.

Most of the theoretical works that have studied the superconducting state of these novel metal hydrides have shown that the strong electron-phonon coupling and the high-energy hydrogen phonon modes play a key role in the superconducting state of these compounds, concluding that the metal hydrides are phonon-mediated strong-coupling superconductors \cite{PhysRevB.96.100502,PhysRevB.91.220507,cor4,cor5,PhysRevLett.114.157004,cor1,cor2,Duan2019,CUI20172526,Flores-Livas2016,Szczniak201730,PhysRevB.93.094525,PhysRevB.93.104526,0022-3719-7-15-015,cor6,BI2019}. Specifically for H$_{3}$S, the compound of interest in this work, experimental results from optical reflectivity (under hydrostatic pressure of 150 GPa) have provided strong evidence of this conventional mechanism. In the same experiment, an unusual strong optical phonon suggests a contribution of electronic degrees of freedom \cite{CarbotteOp,PhysRevB.100.094505}. In several theoretical works it has also been suggested that van Hove singularities in the electronic density of states near the Fermi level (a feature also presented in LaH$_{10}$) play a key role in the high superconducting $T_{c}$ of H$_{3}$S \cite{PhysRevB.93.104526,Bianconi10,Bianconi_2015,italiano,PhysRevB.93.094525,CUI20172526,PhysRevB.99.140501}.

For superconductors, besides $T_{c}$, another important parameter is the isotope effect coefficient. Historically, this parameter has been crucial in elucidate the mechanism responsible for Cooper-pair formation in conventional superconductors. It accounts for the response of the phonon spectrum, the electron-phonon coupling, and the screening electron-electron repulsion to an isotopic mass change, thus giving information on how the lattice dynamics (ion-movement patterns) is involved in the value of $T_{c}$.

For a conventional superconductor, if only phonons are taken into account, the Bardeen-Cooper-Schrieffer theory (BCS)\cite{PhysRev.108.1175} predicts that the $T_{c}$ goes as $M^{-\alpha}$, where $M$ is the isotope mass and $\alpha=0.5$ is the isotope effect coefficient. However, experimental observations have shown that there is not general behavior for $\alpha$. In MgB$_{2}$ \cite{MgB2} and the Rb$_{3}$C$_{60}$ \cite{PhysRevLett.83.404}, for example, $\alpha$ is substantially reduced from the BCS value (0.32 and 0.21 for MgB$_{2}$ and Rb$_{3}$C$_{60}$ respectively). In PdH, an archetypal system for metal hydrides, the isotope coefficient is negative ($\alpha\cong-0.3$) and under pressure it diminishes steadily \cite{PhysRevB.39.4110,VILLACORTES2018371}. For LaH$_{10}$ a value of $\alpha\cong0.47$ at 150 GPa has been observed\cite{La50}, whereas in H$_{3}$S it has a very large value of $\alpha\cong2.37$ at 130 GPa, decreasing it rapidly as pressure increases up to 230 GPa ($\alpha\cong0.29$). While for PdH there are in literature several theoretical and experimental publications studying and discussing the nature of its inverse isotope effect\cite{meromio,VILLACORTES2018371,PhysRevLett.111.177002,PhysRevLett.57.2955,YUSSOUFF1995549}, for H$_{3}$S there are just a few theoretical reports that address this topic\cite{Szczniak201730,cor5,PhysRevB.91.224513,0953-2048-30-4-045011,Harshman_2017,Gor'kov}, but without a proper discussion of the nature of such interesting phenomena or quantitative analysis of $\alpha$.

It is important to mention that almost all the previously mentioned theoretical works are based on the Migdal-Eliashberg theory, which takes into account the detailed shape of the phonon spectrum and the electron-phonon interaction, whereas the electronic density of states $N(\varepsilon)$ is assumed constant in an energy range of the order of the phonon energy, around the Fermi energy. In this case, the isotropic linearized Migdal-Eliashberg Equations (LMEE) are averaged over the Fermi surface, keeping $N(\varepsilon)$ fixed to the Fermi level: $N(0)$ \cite{PhysRevB.13.1416,Eliashberg,ALLEN19831,Bergmann1973}. However, when considering $N(\varepsilon)$ variable (on the scale of phonon energy) it has been shown that both, $T_{c}$ and $\alpha$, are modified, indicating that the nature of the electronic properties on such energy range is crucial for a proper analysis of superconductivity\cite{PhysRevB.42.406,PhysRevB.56.6107}. Therefore, a generalized version of LMEE that includes the energy dependence of $N(\varepsilon)$ has to be adopted \cite{Horsch1977,LIE1978511,PhysRevB.42.406,ALLEN19831,PhysRevB.93.094525,PhysRevB.56.6107}. Such is the case for the A15 family of superconductors and H$_{3}$S, where for the former has been shown that due to the particular shape of $N(\varepsilon)$ the repulsive part of the electron-phonon interaction can be greatly reduced, thereby leading to an additional enhancement of $T_{c}$ \cite{Horsch1977,LIE1978511}, while for the later has been pointed out the importance of the van Hove singularities in the proper understanding of its superconducting state\cite{PhysRevB.93.094525,PhysRevB.93.104526}. 

Here we use the LMEE with ($N(\varepsilon)$) and without ($N(0)$) the energy dependence of the density of states $N(\varepsilon)$ to perform a detailed analysis of $T_{c}$ and the isotope effect. We use a functional derivative approach that relates the changes in the Eliashberg function ($\alpha^{2}F\left(\omega\right)$) and $N(\varepsilon)$, due to the applied pressure, to changes in $T_{c}$. For $\alpha$, we take into account the changes in the electron-electron interaction coming from the isotope mass substitution, in addition to the electron-phonon interaction. Then, the isotope coefficient is splitted into two contributions, each of them to their corresponding interaction \cite{meromio,VILLACORTES2018371,PhysRevB.30.5019}. This formalism has successfully explained the inverse isotope effect in PdH at ambient pressure\cite{meromio}, as well as its behavior under pressure\cite{VILLACORTES2018371}. For the electron-phonon contribution we use a generalization of the work of Rainer and Culleto on the differential isotope effect coefficient of $\beta\left(\omega\right)$ that includes energy dependence in $N(\varepsilon)$ \cite{PhysRevB.19.2540,PhysRevB.42.406}. The differential coefficient $\beta\left(\omega\right)$ helps us understand how energy dependence in $N(\varepsilon)$ could modify $\beta\left(\omega\right)$ and, by implication, $\alpha$. In addition, the effects of the van Hove singularity to the electron-electron interaction are analyzed, as well as their contribution to the isotope effect. Finally zero point energy (ZPE) is taken into account through the quasi-harmonic approximation (QHA)\cite{10.2138/rmg.2010.71.3,PhysRevB.99.214504}. 

The paper is organized as follows. The theory and the basic equations that support our method are presented in Section 2. The Section 3 contains the technical details. In Sections 4 and 5 we present our results related to the analysis of the isotope effect. Last, our conclusions are presented in Section 6.

\section{Theory and basic equations}

For a compound with several atoms, the isotope coefficient is defined as $\alpha_{i}=-d\ln T_{c}/d\ln M_{i}$, where $M_{i}$ is the mass of the $i$-th atom in the compound, and $T_{c}$ the critical temperature. The total isotope coefficient is given by the sum of the partial ones, namely $\alpha^{total}=\sum_{i}\alpha_{i}$. 

According to Migdal-Eliashberg theory, $T_{c}$ depends on both, the electron-phonon interaction and the electron-electron repulsion. Then, taking that into account, the partial isotope coefficients can be splitted into two contributions\cite{meromio,VILLACORTES2018371,PhysRevB.30.5019}: one that comes from the change in $T_{c}$ due to changes in the electron-phonon interaction, namely $\alpha_{i}^{el-ph}$, while the second one comes from the change in $T_{c}$ due to changes in the electron-electron interaction, $\alpha_{i}^{el-el}$. The partial isotope coefficient is then given by 
\begin{equation}
\alpha_{i}=\alpha_{i}^{el-ph}+\alpha_{i}^{el-el},\label{eq:Totalalfa}
\end{equation}
and the total change in the critical temperature, $\Delta T_{c}$, due to the isotope mass substitution is calculated from
\begin{equation}
\Delta T_{c}^{total}=\Delta T_{c}^{el-ph}+\Delta T_{c}^{el-el}.
\end{equation}

For the study of mass dependence on the electron-phonon interaction and the changes on $T_{c}$ originated by it, Rainer and Culetto \cite{PhysRevB.19.2540} have shown that $\Delta T_{c}^{el-ph}$ and their corresponding total isotope coefficient $\alpha^{el-ph}$ can be calculated as
\begin{equation}
\Delta T_{c}^{el-ph}=\int_{0}^{\infty}d\omega\frac{\delta T_{c}}{\delta\alpha^{2}F\left(\omega\right)}\Delta\alpha^{2}F\left(\omega\right),\label{eq:tc-el-ph}
\end{equation}
and 
\begin{equation}
\alpha^{el-ph}=\int_{0}^{\infty}d\omega\beta^{el-ph}\left(\omega\right),\label{eq:IntAlfa}
\end{equation}
where 
\begin{equation}
\beta^{el-ph}\left(\omega\right)\equiv-\frac{\delta T_{c}}{\delta\alpha^{2}F\left(\omega\right)}\frac{\Delta\alpha^{2}F\left(\omega\right)}{T_{c}\Delta\ln M},\label{eq:dif_alfa}
\end{equation}
is the differential isotope coefficient, that can be expressed by an equivalent expression
\begin{equation}
\beta^{el-ph}\left(\omega\right)\equiv R\left(\omega\right)\alpha^{2}F\left(\omega\right),
\end{equation}
where $R\left(\omega\right)$ is given by 
\begin{equation}
R\left(\omega\right)=\frac{d}{d\omega}\left[\frac{\omega}{2T_{c}}\frac{\delta T_{c}}{\delta\alpha^{2}F\left(\omega\right)}\right].\label{eq:we}
\end{equation}
In the previous equations, $\alpha^{2}F\left(\omega\right)$ is the Eliashberg function, defined as

\begin{eqnarray}
\alpha^{2}F\left(\omega\right)= &  & \frac{1}{N\left(0\right)}\sum_{nm}\sum_{\vec{q}\nu}\delta\left(\omega-\omega_{\vec{q}\nu}\right)\sum_{\vec{k}}\left|g_{\vec{k}+\vec{q},\vec{k}}^{\vec{q}\nu,nm}\right|^{2}\label{eq:a2f-def}\\
 &  & \times\delta\left(\varepsilon_{\vec{k}+\vec{q},m}-\varepsilon_{F}\right)\delta\left(\varepsilon_{\vec{k},n}-\varepsilon_{F}\right),\nonumber 
\end{eqnarray}
where $g_{\vec{k}+\vec{q},\vec{k}}^{\vec{q}\nu,nm}$ are the matrix elements of the electron-phonon interaction, $\varepsilon_{\vec{k}+\vec{q},m}$ and $\varepsilon_{\vec{k},n}$ are one-electron band energies, with band index $m$ and $n$, with vectors $\vec{k}+\vec{q}$ and $\vec{k}$, respectively, and $\omega_{\vec{q}\nu}$ is the phonon frequency for mode $\nu$ at wave-vector $\vec{q}$. The $T_{c}$ functional derivative respect to the Eliashberg function\cite{Bergmann1973,Daams1979}, $\delta T_{c}/\delta\alpha^{2}F\left(\omega\right)$, can be calculated from the solution of the LMEE as discussed at the end of this section.

The effects of electron-electron contribution on the isotope coefficient, $\alpha^{el-el}$, can be extracted from the Coulomb repulsion parameter $\mu^{*}$ which also depends on the phonon frequency, and therefore it can be modified due to isotope substitution\cite{LEAVENS19741329}. $\mu^{*}$ represents an effective potential that gives a measure on how retardation effects (due to the electron-phonon interaction) influence the bare Coulomb potential. The partial isotope coefficient for the electron-electron interaction is given by 
\begin{equation}
\alpha_{i}^{el-el}=-\frac{\Delta T_{c}^{el-el}}{T_{c}\Delta\ln M_{i}},
\end{equation}
where the change in $T_{c}$ due to the electron-electron interaction
is expressed as\cite{meromio} 
\begin{equation}
\Delta T_{c}^{el-el}=\frac{\partial T_{c}}{\partial\mu^{*}}\left(\mu_{D_{3}S}^{*}-\mu_{H_{3}S}^{*}\right).\label{eq:tcel-el}
\end{equation}

Then, according to Eqs. \ref{eq:Totalalfa}-\ref{eq:tcel-el}, in order to calculate $\alpha_{i}$ it is necessary to know first $\mu^{*}$ and $\delta T_{c}/\delta\alpha^{2}F\left(\omega\right)$. $\mu^{*}$ can be obtained by solving the LMEE valid at $T_{c}$ once $\alpha^{2}F\left(\omega\right)$ is known and then, from the solution of the LMEE, the functional derivative, $\delta T_{c}/\delta\alpha^{2}F\left(\omega\right)$, can be calculated by the Bergmann and Rainer formalism\cite{Bergmann1973}. 

When $N(\varepsilon)$ is taking into account in the LMEE, $T_{c}$ becomes a functional of it too. Furthermore, a change $\Delta N(\varepsilon)$ in the electronic density of states results in a change in the critical temperature given by 
\begin{equation}
\Delta T_{c}^{\Delta N(\varepsilon)}=\int_{0}^{\infty}d\varepsilon\frac{\delta T_{c}}{\delta N(\varepsilon)}\Delta N(\varepsilon),\label{eq:deltados}
\end{equation}
where $\delta T_{c}/\delta N(\varepsilon)$ is the functional derivative of $T_{c}$ respect to $N(\varepsilon)$.
\begin{figure}
\includegraphics[width=8.4cm]{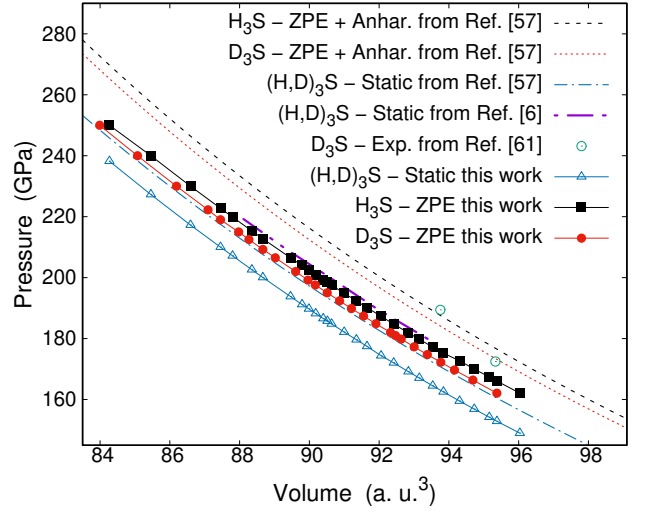}
\caption{\label{fig:PV}Calculated state equation for $\textrm{H}_{3}\textrm{S}$ and $\textrm{D}_{3}\textrm{S}$ with (ZPE) and without (static) zero-point energy contribution. Results from Duan et al. \cite{10.1038/srep06968} and Errea et al. \cite{CLm3m} (where anharmonic effects were included) are also shown.}
\end{figure}
In the presented research, we have used both LMEE schemes: the generalized one, that includes the energy dependence of $N(\varepsilon)$\cite{Horsch1977,LIE1978511,PhysRevB.42.406,ALLEN19831}, and the standard one, that considers $N(\varepsilon)$ as a constant: $N(0)$. For the energy-dependent case, the superconducting gap equations on the imaginary axis are
\begin{equation}
\tilde{\Delta}_{n}=\pi T\sum_{m}\left[\lambda_{nm}-\mu^{*}\right]\frac{\tilde{\Delta}_{m}}{\left|\tilde{\omega}_{m}\right|}\tilde{N}(\left|\tilde{\omega}_{m}\right|),
\end{equation}
and 
\begin{equation}
\tilde{\omega}_{n}=\omega_{n}+\pi T\sum_{m}\lambda_{nm}\textrm{sig}(\omega_{m})\tilde{N}(\left|\tilde{\omega}_{m}\right|),
\end{equation}
where $T$ is the temperature and $\omega_{n}$ are the Matsubara frequencies, expressed as $i\omega_{n}=i\pi T\left(2n-1\right)$, with $n=0,\pm1,\pm2,\ldots$. Under this framework, the coupling parameter $\lambda_{nm}$ is defined as 
\begin{equation}
\lambda_{nm}=2\int_{0}^{\infty}d\omega\frac{\omega\alpha^{2}F\left(\omega\right)}{\omega^{2}+\left(\omega_{m}-\omega_{n}\right)^{2}},\label{eq:el-ph-parameter}
\end{equation}
with $\lambda_{nn}$ as the electron-phonon parameter, and 
\begin{equation}
\tilde{N}(\left|\tilde{\omega}_{n}\right|)=\frac{1}{\pi}\int_{-\infty}^{\infty}d\varepsilon\frac{N\left(\varepsilon\right)}{N\left(0\right)}\frac{\left|\tilde{\omega}_{n}\right|}{\left|\tilde{\omega}_{n}\right|^{2}+\varepsilon^{2}}.
\end{equation}
Introducing the breaking parameter $\rho$ (that becomes zero at $T_{c}$), and with the definition $\bar{\Delta}_{n}\equiv\tilde{\Delta}_{n}/\left[\left|\tilde{\omega}_{n}\right|+\rho\right]$ we get the LMEE valid at $T_{c}$ expressed as: 
\begin{equation}
\rho\bar{\Delta}_{n}=\sum_{m}K_{nm}\bar{\Delta}_{m},\label{eq:LMEE}
\end{equation}
with 
\begin{equation}
K_{nm}=\pi T_{c}\left[\lambda_{nm}-\mu^{*}\right]\tilde{N}(\left|\tilde{\omega}_{m}\right|)-\delta_{nm}\left|\tilde{\omega}_{m}\right|.
\end{equation}
For the case of $N(\varepsilon)=N(0)=$const. there is a well known set of equations similar to Eq. (\ref{eq:LMEE}), but in this case the kernel $K_{nm}$ is given by\cite{Bergmann1973,VILLACORTES2018371,meromio,Daams1979}
\begin{equation}
K_{nm}=\pi T_{c}\left[\lambda_{nm}-\mu^{*}-\delta_{nm}\frac{\left|\tilde{\omega}_{m}\right|}{\pi T_{c}}\right],\label{eq:NDEN}
\end{equation}
with 
\begin{equation}
\tilde{\omega}_{n}=\omega_{n}+\pi T\sum_{m}\lambda_{nm}\textrm{sig}(\omega_{m}).
\end{equation}
For both versions of the LMEE, the functional derivative of $T_{c}$
respect to the Eliashberg function is given by 
\begin{equation}
\frac{\delta T_{c}}{\delta\alpha^{2}F\left(\omega\right)}=-\left.\left(\frac{d\rho}{dT}\right)\right|_{T_{c}}\frac{\delta\rho}{\delta\alpha^{2}F\left(\omega\right)},\label{eq:dev_a2f}
\end{equation}
while the functional derivative of $T_{c}$ respect to $N(\varepsilon)$
(when $N(\varepsilon)$ varies) is given by (see Appendix \ref{sec:The-functional-derivatives}):
\begin{equation}
\frac{\delta T_{c}}{\delta N(\varepsilon)}=-\left.\left(\frac{d\rho}{dT}\right)\right|_{T_{c}}\frac{\delta\rho}{\delta N(\varepsilon)}.\label{eq:dev_ne}
\end{equation}

\section{Numerical Parameters}
\begin{figure}
\includegraphics[width=8.4cm]{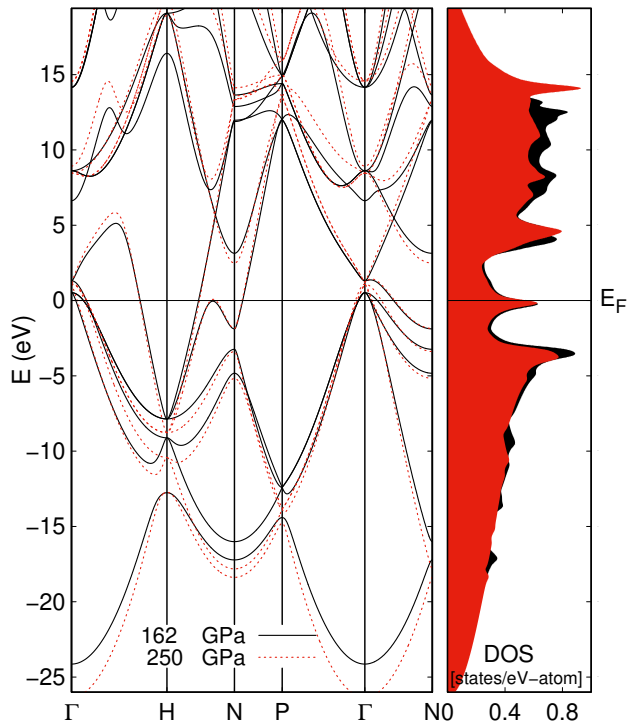}\caption{\label{fig:bands}Calculated electronic band dispersion and electronic
density of states for H$_{3}$S at 162 and 250 GPa.}
\end{figure} 
In order to study the isotope effect, and therefore the superconducting state, the electronic structure, the phonon dispersion and the electron-phonon coupling properties were obtained. We calculated the ground state properties within the framework of Density Functional Theory (DFT)\cite{PhysRev.140.A1133}, while the lattice dynamics and coupling properties within the Density Functional Perturbation Theory (DFPT)\cite{0953-8984-21-39-395502,RevModPhys.73.515}, both implemented in the QUANTUM ESPRESSO suit code \cite{0953-8984-21-39-395502}. The studied properties were obtained for the $Im\bar{3}m$ crystal structure in a range from 162 to 250 GPa\cite{eina,CLm3m,doi:10.1002/chem.201705321}. The calculations were performed with a 80 Ry cutoff for the plane-wave basis and a $32\times32\times32$ $k$-point mesh. We use the scalar relativistic pseudo potentials of Perdew and Zunger (LDA) \cite{PhysRevB.23.5048} since LDA has been used previously, showing a good performance\cite{Flores-Livas2016,PhysRevB.93.020508,PhysRevB.93.224513}.

\begin{figure}
\includegraphics[width=8.4cm]{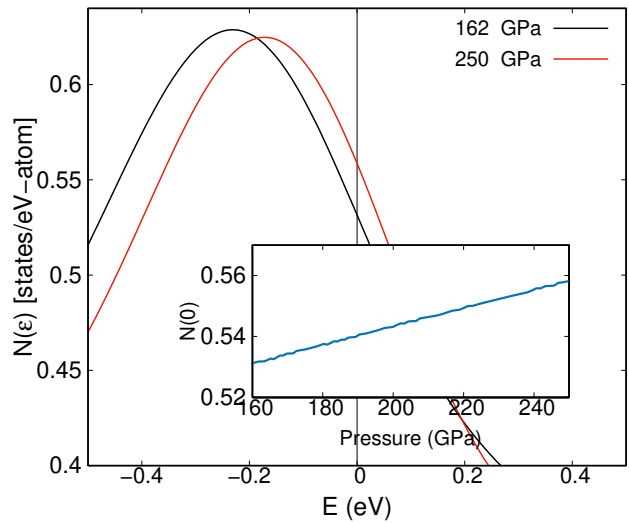}
\caption{\label{fig:It-shows-thedos}$N(\varepsilon)$ behavior near the Fermi level for H$_{3}$S at 165 and 250 GPa. The inset shows the density of states at $E_{F}$ ($N(0)$) in a broad interval of pressures.}
\end{figure}

Dynamical matrices were calculated on a $8\times8\times8$ $q$-point mesh, while for the calculation of electron-phonon matrix elements a higher mesh of $32\times32\times32$ $k$-point was required. Finally, the phonon density of states and Eliashberg function were obtained from a Fourier interpolated denser mesh of $72\times72\times72$ $q$-point.

Corrections due to quantum fluctuations at zero temperature are estimated through the quasi-harmonic approximation (QHA)\cite{10.2138/rmg.2010.71.3,PhysRevB.99.214504} using the calculated phonon density of states (PDOS). Within this approximation, the phonon contribution to the ground-state energy is taken into account and a new structural optimization can be performed. Thus, the electronic structure, lattice dynamics and electron-phonon properties calculated with these lattice parameters include ZPE corrections.

Finally, to solve the LMEE a cut-off frequency $\omega_{c}$ is defined in terms of the maximum phonon frequency $\omega_{ph}$, in order to cut the sum over the Matsubara frequencies. We perform our calculations with a cut-off of $\omega_{c}=1000\times\omega_{ph}$. Once $\omega_{c}$ is set, the only adjustable parameter left is $\mu^{*}$, which is fitted at each pressure of interest.

\section{Ground state, lattice dynamics, and coupling properties}
\begin{figure*}
\includegraphics[width=17cm]{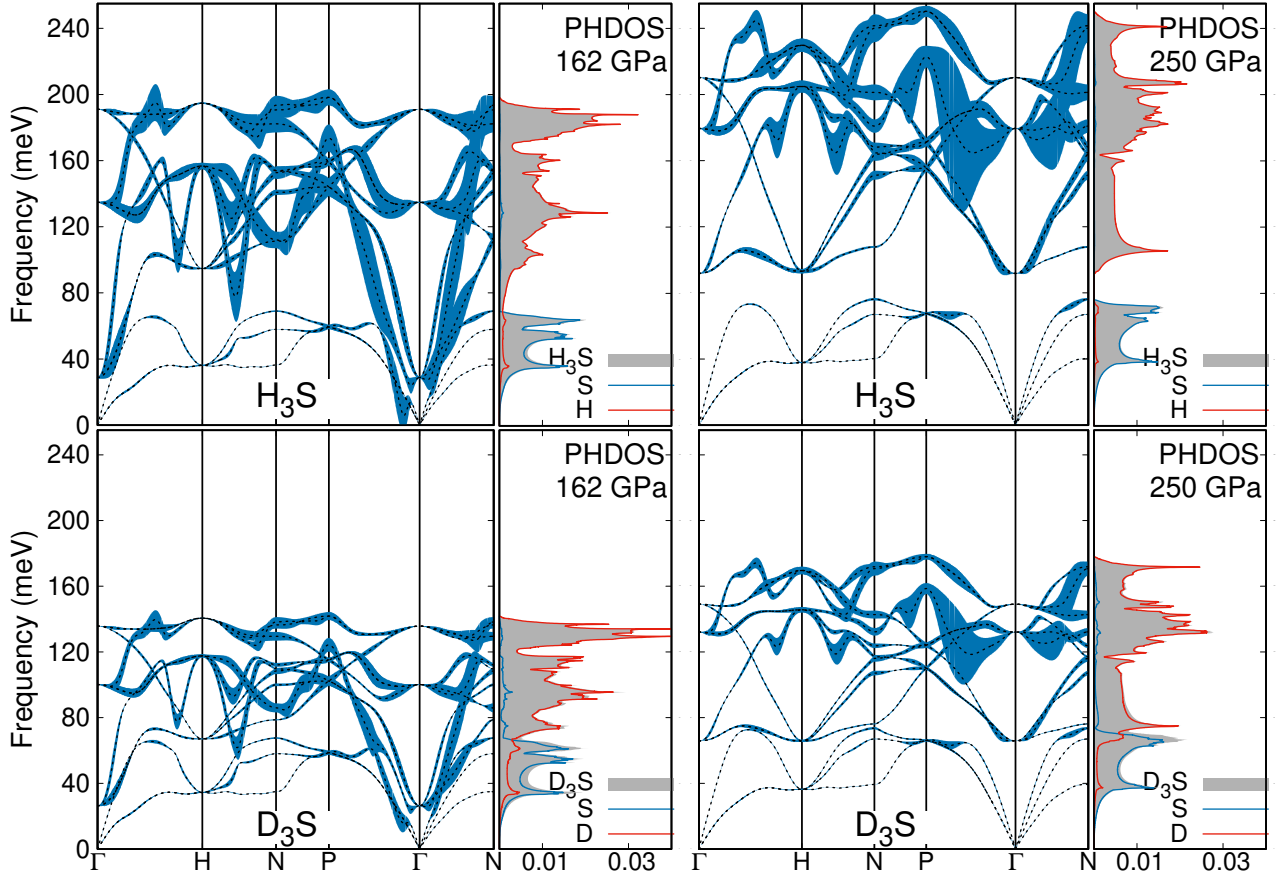}\caption{\label{fig:modos}Phonon dispersion, total and projected phonon density of states calculated for H$_{3}$S (top) and D$_{3}$S (bottom) at 162 and 250 GPa. The vertical lines at $\omega_{\vec{q}}$ represent the associated phonon linewidhts.}
\end{figure*}
Crystal lattice parameters were optimized within QHA at pressures (from 162 to 250 GPa) where experimental values of $T_{c}$ have been measured. In Fig. \ref{fig:PV} we present the equation of state for H$_{3}$S and D$_{3}$S with (ZPE) and without (static) considering the zero-point motion contribution to the energy. The correction from the ZPE contribution to the pressure is smaller in D$_{3}$S than for H$_{3}$S due to the smaller phonon energy scale in the former. For comparison, the equations of state previously reported\cite{CLm3m,10.1038/srep06968,eina} are also shown. In Fig. \ref{fig:bands} we show the electronic band structure and the density of states (which include ZPE corrections) for H$_{3}$S at some pressures of interest. The present calculations are in good agreement with previous band structure calculations \cite{PhysRevB.91.184511,10.1038/srep06968,Bianconi10,italiano}. At 162 GPa the band structure consists of six valence bands, four of them are filled, leaving some holes at $\mathsf{\varGamma}$, while the fifth and sixth are broad bands roughly half-filled. The tops of the hole-like bands near the $\mathsf{\varGamma}$ point are above the Fermi level ($E_{F}$) and they move slowly far away from $E_{F}$ as the applied pressure increases. A detailed analysis of the behavior of the band structure as a function of pressure has been done previously by Jarlborg et al.\cite{italiano}. As can be seen from the DOS, there is a van Hove singularity close to $N(0)$. In order to observe it closer, we show in Fig \ref{fig:It-shows-thedos} the behavior of $N(\varepsilon)$ near the Fermi level, where it can be seen how the narrow peak moves towards $E_{F}$ as pressure increases. The inset shows how $N(0)$ increases as pressure rises up on H$_{3}$S.

In Fig. \ref{fig:modos} we show the phonon spectrum and the phonon density of states (PHDOS) for both compounds, H$_{3}$S and D$_{3}$S, at 162 and 250 GPa, including ZPE. In general, for both of them, the observed pressure effect is a hardening of the phonon frequencies as pressure arises from 162 to 250 GPa, specially for the H(D) related modes, while the modes corresponding to S remain almost unchanged. In particular, for H$_{3}$S at 250 GPa it is observed a clear separation into H modes at high energy (optical branches) and S modes (acoustic branch) below 80 meV, in agreement with previous works\cite{10.1038/srep06968,Flores-Livas2016,PhysRevB.91.220507,PhysRevLett.111.177002,PhysRevB.93.094525,PhysRevB.91.224513,cor1}. When the H ions are substituted by their isotopes D, there is a strong re-normalization of the optical phonon modes to lower energies. As a result, the electron-phonon and superconducting properties are going to be affected due to isotope mass substitution. To gain more insight, Fig. \ref{fig:modos} also shows phonon linewidths of the $\vec{q}\nu$ phonon mode, $\gamma_{\vec{q}\nu}$, arising from the electron-phonon interaction given by\cite{PhysRevB.6.2577,PhysRevB.9.4733}
\begin{equation}
\gamma_{\vec{q}\nu}=2\pi\omega_{\vec{q}\nu}\sum_{\vec{k}}^{nm}\left|g_{\vec{k}+\vec{q},\vec{k}}^{\vec{q}\nu,nm}\right|^{2}\delta\left(\varepsilon_{\vec{k}+\vec{q},m}-\varepsilon_{F}\right)\delta\left(\varepsilon_{\vec{k},n}-\varepsilon_{F}\right).
\end{equation}
The phonon linewidths of the H and D vibrations are fairly uniform throughout the spectrum, showing larger values for the H vibrations than the D ones, and increasing as pressure arises for both systems.

\begin{figure}
\includegraphics[width=8.4cm]{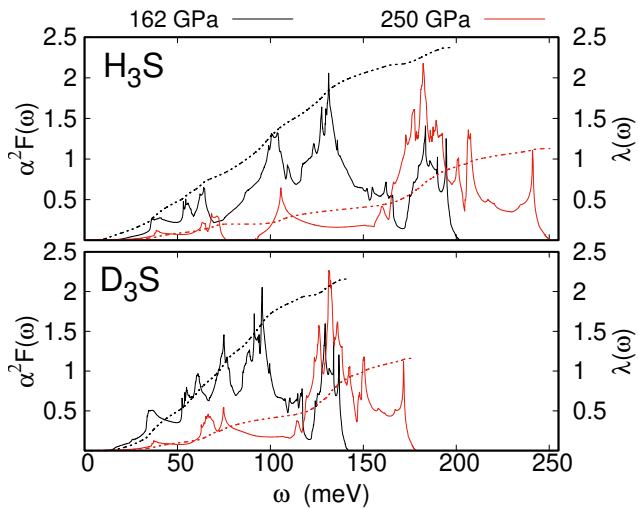}
\caption{\label{fig:A2f_zp_pressure}Eliashberg function (solid lines) and electron-phonon coupling constant (dashed lines) for H$_{3}$S and D$_{3}$S at 162 and 250 GPa.}
\end{figure}

\begin{figure}
\includegraphics[width=8.4cm]{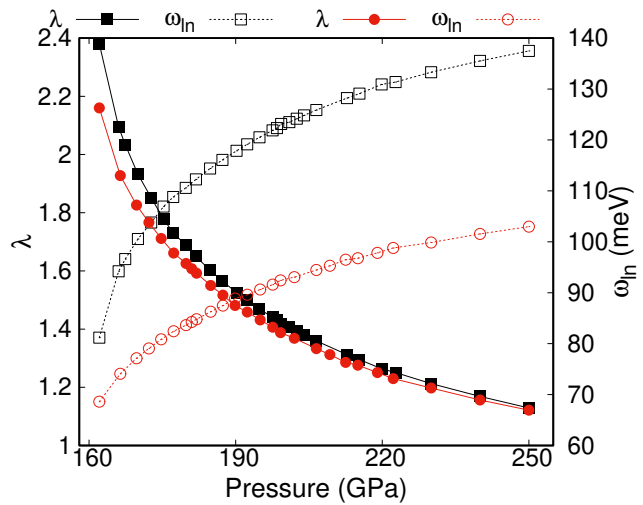}
\caption{\label{fig:lamb-wln}Electron-phonon coupling constant ($\lambda$), and the Allen-Dynes characteristic phonon frequency ($\omega_{ln}$), for H$_{3}$S (squares) and D$_{3}$S (circles) as a function of the applied pressure.}
\end{figure}

In Fig. \ref{fig:A2f_zp_pressure} 
 we show the Eliashberg function as well as the electron-phonon parameter, $\lambda$, for both compounds at 162 and 250 GPa. For H$_{3}$S, the maximum phonon frequency of the spectrum corresponding to H-modes shifts to a higher frequency region (approx. 52 meV) as the pressure arises from 162 GPa to 250 GPa. For D$_{3}$S a similar effect is observed, however, the shift is smaller (approx. 36 meV). The observed shift has an impact on the electron-phonon coupling constant as well (defined by $\lambda_{nn}$ in Eq. \ref{eq:el-ph-parameter}), since it decreases from 2.38 to 1.13 for H$_{3}$S and from 2.16 to 1.12 for D$_{3}$S as pressure goes from 162 to 250 GPa. In Fig. \ref{fig:lamb-wln}
 the evolution of $\lambda$, for H$_{3}$S and D$_{3}$S, as pressure increases is shown, as well as the Allen-Dynes characteristic phonon frequency\cite{PhysRevB.12.905} defined by 
\begin{equation}
\omega_{ln}=\exp\left\{ \frac{2}{\lambda}\int_{0}^{\infty}d\omega\frac{\alpha^{2}F\left(\omega\right)}{\omega}\ln\omega\right\} .
\end{equation}
For H$_{3}$S the calculated $\omega_{ln}$ value raises from 81.18 to 137.49 meV going from 162 to 250 GPa, which is consistent with previous works\cite{10.1038/srep06968,Flores-Livas2016,PhysRevB.91.220507,PhysRevLett.111.177002,PhysRevB.93.094525,PhysRevB.91.224513,cor1}, while for D$_{3}$S goes from 68.61 to 102.68 meV for the same pressure values. As it is expected, both parameters ($\lambda$ and $\omega_{ln}$) are affected due to the isotope mass. The shift to lower frequencies for $\omega_{ln}$, due to the renormalization of the phonon spectrum, increases as pressure arises, going from 12.57 meV at 162 GPa until 34.81 meV at 250 GPa. For $\lambda$, the effect is less dramatical, showing a 10\% reduction at 162 GPa, and reducing such difference to less than 1\% at 250 GPa.

\section{The Isotope Effect in (H,D)$_{3}$S}

In order to know $\alpha$ it is necessary, first, to know the values of $T_{c}$ for H$_{3}$S and D$_{3}$S at similar pressures. However, it is worth to note that the experimental available data for both compounds are very scattered and do not correspond to similar applied pressures. Therefore, first we must get $T_{c}(P)$ from a theoretical method that it is described below. This method\cite{VILLACORTES2018371,ivan} assumes previous knowledge of the critical temperature at some starting pressure, $P_{0}$: $T_{c}\left(P_{0}\right)$, which is the only required input parameter.

\subsection*{The critical temperature}

\begin{figure}
\includegraphics[width=8.4cm]{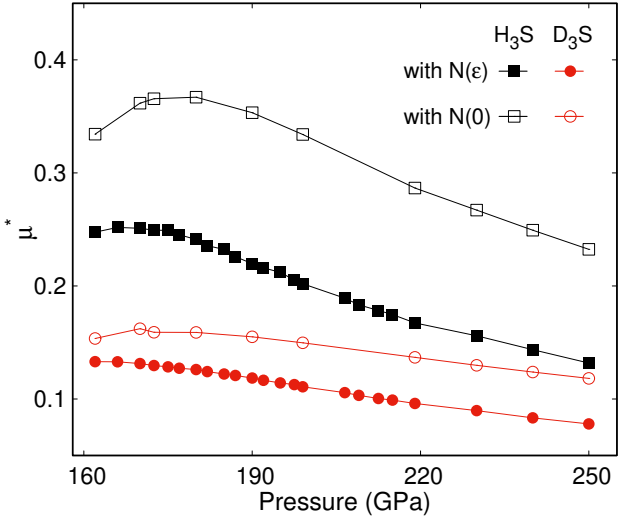}
\caption{\label{fig:mus}$\mu^{*}$ for H$_{3}$S and D$_{3}S$ at several
pressures.}
\end{figure}

As was already mentioned, in order to calculate $T_{c}\left(P\right)$, we have to start at an initial pressure where the critical temperature is known, that is $T_{c}\left(P_{0}\right)$. From it, and with the calculated $\alpha^{2}F\left(\omega,P_{0}\right)$, $\mu^{*}\left(P_{0}\right)$ can be obtained by fitting it to $T_{c}\left(P_{0}\right)$ through LMEE (see Fig. \ref{fig:mus}).
 Then by Bergmann and Rainer\cite{Bergmann1973,LIE1978511} formalism described in Sec. II, we calculated the functional derivative of $T_{c}$ respect to $\alpha^{2}F\left(\omega\right)$ and $N(\varepsilon)$ (Eq. (\ref{eq:dev_a2f}) and Eq. (\ref{eq:dev_ne})). These functional derivatives allow to relate the observed changes on $\alpha^{2}F\left(\omega\right)$ and $N(\varepsilon)$ by a pressure variation (from $P_{0}$ to $P_{i}$) to changes in the critical temperature, $\Delta T_{c}(P_{0},P_{i})$. The resulting change of $T_{c}$ from $P_{0}$ to $P_{i}$ is given by\cite{LIE1978511,Bergmann1973,VILLACORTES2018371}
\begin{equation}
\Delta T_{c}(P_{0},P_{i})=\Delta T_{c}^{el-ph}+\Delta T_{c}^{N\left(\varepsilon\right)},
\end{equation}
where $\Delta T_{c}^{el-ph}$ and $\Delta T_{c}^{N\left(\varepsilon\right)}$ are given by Eq. (\ref{eq:tc-el-ph}) and Eq. (\ref{eq:deltados}) respectively. Then, the critical temperature for the system under pressure $P_{i}$ is given by
\begin{equation}
T_{c}(P_{i})=T_{c}(P_{0})+\Delta T_{c}(P_{0},P_{i}).
\end{equation}

\begin{figure}
\includegraphics[width=8.4cm]{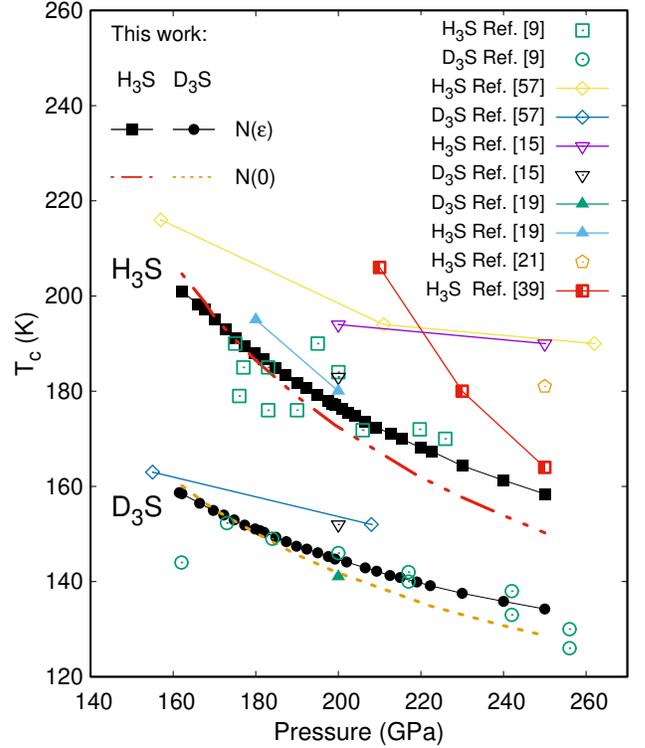}
\caption{\label{fig:TC_P}Calculated $T_{c}(P)$ for both compounds, H$_{3}$S and D$_{3}$S, within the Migdal-Eliashberg formalism with $N(\varepsilon)$ and $N(0)$. For comparison it is shown the experimental data from Drozdov et al. \cite{203} for H$_{3}$S (open squares) and D$_{3}$S (open circles), as well as previously reported $T_{c}$ calculations\cite{PhysRevLett.114.157004,CLm3m,Flores-Livas2016,PhysRevB.93.094525,PhysRevB.91.224513}.
}
\end{figure}

In Fig. \ref{fig:TC_P} 
 it is presented $T_{c}$ as a function of pressure, in the range of interest for H$_{3}$S and D$_{3}$S, calculated with the method described above with ($N(\varepsilon)$) and without ($N(0)$) the energy dependence of the density of states close to the Fermi level in LMEE. We use, for D$_{3}$S, $T_{c}=154$ K at 172 GPa as our starting data. In the case of H$_{3}$S, we took as our starting temperature $T_{c}=193$ K at 172 GPa. It can be observed that our calculated $T_{c}$ for D$_{3}$S is in excellent agreement with the experimental values (open circles) in the whole pressure interval (162-250 GPa) for both schemes: $N(\varepsilon)$ and $N(0)$. For H$_{3}$S the calculated $T_{c}$ for both schemes ($N(\varepsilon)$ and $N(0)$) follow nicely the experimentally observed reduction of $T_{c}$ as the applied pressure increases. It is important to mention that, for both systems, we observe with the $N(0)$ scheme lower $T_{c}$ values than the ones calculated with the $N(\varepsilon)$ scheme, moving slightly away from the experimental data. Although such differences are as large as 8 K at the high-pressure regime (around 250 GPa), they do not modify the general discussion of our findings neither the better agreement with experimental data than other theoretical
calculations reported previously in literature\cite{10.1038/srep06968,PhysRevB.93.094525,PhysRevB.91.224513,doi:10.7566/JPSJ.87.124711,PhysRevLett.114.157004,CLm3m}. 

\begin{figure}
\includegraphics[width=8.4cm]{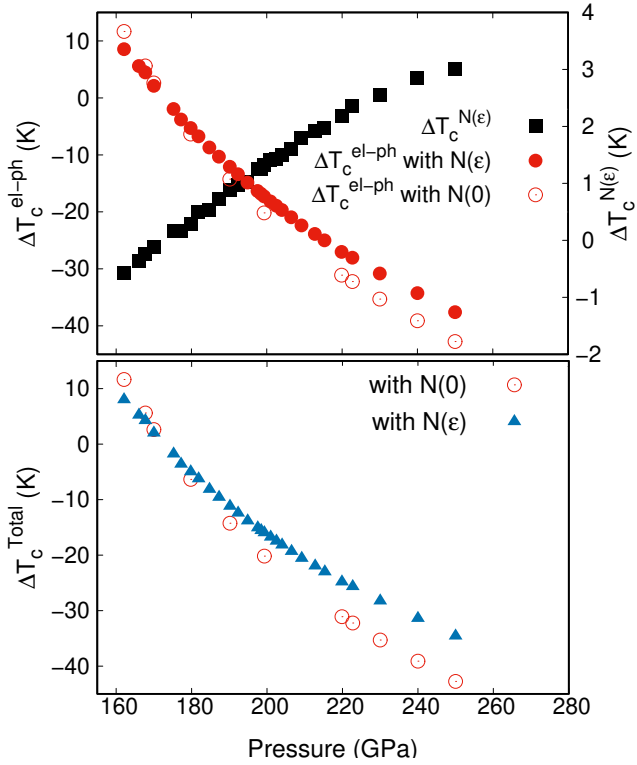}
\caption{\label{fig:deltasTC}The electron-phonon and $N(\varepsilon)$ contributions to the change of critical temperature, as well as the total $\Delta T_{c}$ as a function of pressure, calculated within the Migdal-Eliashberg formalism with ($N(\varepsilon)$) and without ($N(0)$) the energy dependence on the density of states close to $E_{F}$ in the LMEE.}
\end{figure}

In order to understand the evolution of $T_{c}$ as a function of
pressure, we present on Fig. \ref{fig:deltasTC}
 the calculated $\Delta T_{c}^{el-ph}\left(P\right)$, $\Delta T_{c}^{N\left(\varepsilon\right)}\left(P\right)$ and $\Delta T_{c}\left(P\right)$, taking 172 GPa as the reference pressure, $P_{0}$. We found that $\Delta T_{c}^{el-ph}$ decreases as a function of pressure, on both schemes in the LMEE, with $N(\varepsilon)$ and $N(0)$. The pressure effects are slightly stronger for the $N(0)$ scheme at the high-pressure regime. For $\Delta T_{c}^{N\left(\varepsilon\right)}$, on the contrary, it is observed an increase as a function of pressure, although at a minor scale than the electron-phonon contribution. Then, by putting together the previous information on $\Delta T_{c}^{Total}$, it shows that the electron-phonon are the main responsible of the decrease of $T_{c}$ as a function of pressure. In addition, by taking into account $\Delta T_{c}^{N\left(\varepsilon\right)}$ the agreement with the experiments is improved, by compensating, on a minor scale, the strong reduction induced by the electron-phonon term.

\subsection*{The isotope effect coefficient}

\begin{figure}
\includegraphics[width=8.4cm]{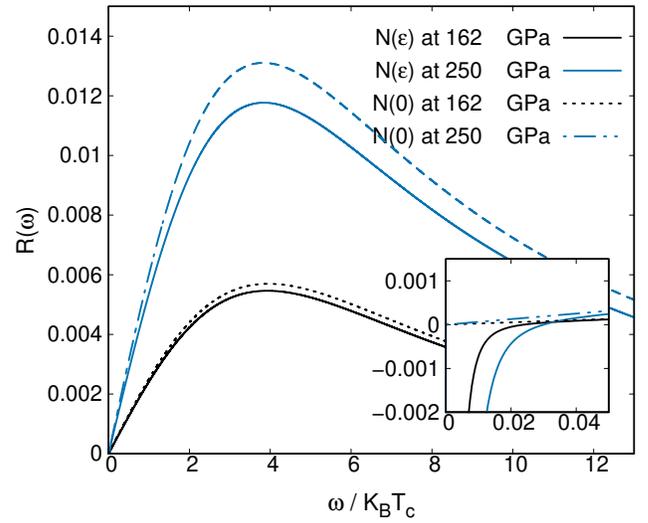}
\caption{\label{fig:rcu}Weighting function $R\left(\omega\right)$ calculated for H$_{3}$S at two specific pressures: 162 and 250 GPa. The full-lines correspond to the $N\left(\varepsilon\right)$ cases (energy dependent) and the dotted ones correspond to $N\left(0\right)$ (constant). The inset shows the low-frequency region where $R\left(\omega\right)$ becomes negative for $N\left(\varepsilon\right)$.}
\end{figure}

\begin{figure}
\includegraphics[width=8.4cm]{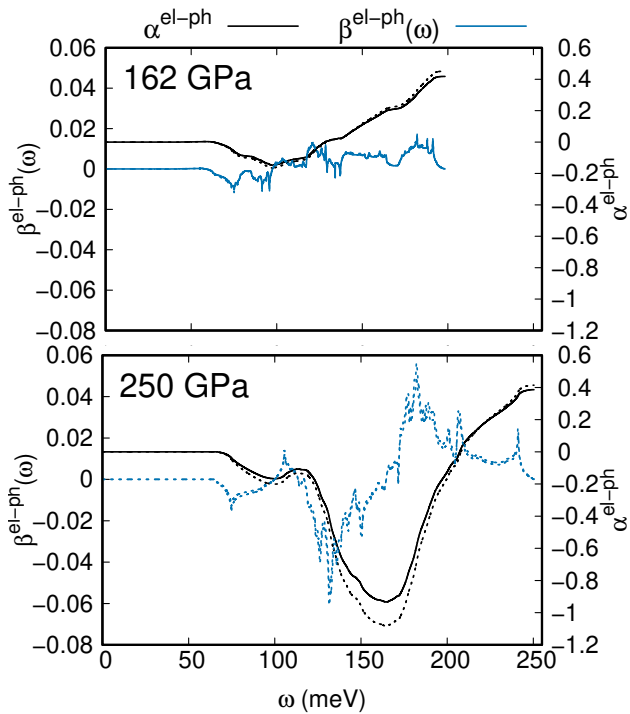}
\caption{\label{fig:alfa_modos}$\beta^{el-ph}\left(\omega\right)$ and partial integrated $\alpha^{el-ph}$ at 162 and 250 GPa. The full lines correspond to the $N(\varepsilon)$ case and the dotted ones to $N(0)$.}
\end{figure}

In Fig. \ref{fig:rcu} we show the calculated weighting function $R\left(\omega\right)$, Eq. (\ref{eq:we}), for the differential isotope effect $\beta^{el-ph}\left(\omega\right)$ at specific pressures. This function determines the contribution to the total isotope effect from the electron-phonon interaction around a frequency $\omega$ due to changes in the isotope mass (the substitution of hydrogen by deuterium). The calculations were performed taking into account both versions of the LMEE, with $N(0)$ (constant) and $N(\varepsilon)$ (energy-dependent) density of states. For the $N(0)$ case, it shows a single broad maximum around a frequency of $4k_{B}T_{c}$, and goes to zero like $\omega$ for $\omega\rightarrow0$ and like $1/\omega$ for $\omega\rightarrow\infty$. Thus, the high-energy phonons are less effective in $\alpha^{el-ph}$ than those at the maximum around $4k_{B}T_{c}$ (68 and 56 meV at 162 and 250 GPa). Previous publication have reported important modifications on the $R\left(\omega\right)$ function\cite{PhysRevB.42.406} when $N\left(\varepsilon\right)$ is considering instead of $N\left(0\right)$ into the calculations , a situation that we also observe from our results. In particular, the H$_{3}$S DOS presents a van Hove singularity near the Fermi level. So, by using $N\left(\varepsilon\right)$, the values of $R\left(\omega\right)$ are reduced on the whole range of frequencies, in comparison with the $N\left(0\right)$ case, becoming negative as $\omega$ approaches to zero. This behavior indicates that $\alpha^{el-ph}$ for $N\left(\varepsilon\right)$ would be smaller than for $N\left(0\right)$, and even negative for the low-energy phonons region. 
In Fig. \ref{fig:alfa_modos} we show the differential isotope coefficient $\beta^{el-ph}\left(\omega\right)$ and the partial integrated value $\alpha^{el-ph}$ (Eqs. (\ref{eq:dif_alfa}) and (\ref{eq:IntAlfa}), respectively). $\beta^{el-ph}\left(\omega\right)$ presents negative values at a broad energy range of approx 47 meV (from 50 meV to 97 meV) at 162 GPa, which is expanded to 99 meV (from 65 meV to 164 meV) at 250 GPa. Due to this behavior, $\alpha^{el-ph}$ shows also a broad region where the partial isotope effect is negative with peak values of -0.15(-0.16) at 97 meV for 162 GPa and -0.93(-1.08) at 164 meV for 250 GPa when $N(\varepsilon)$($N(0)$) is taking into account. This inverse isotope effect (negative values) is associated to an increase of $T_{c}$. However, as the integration is performed in the whole energy range, $\alpha^{el-ph}$ becomes positive, reaching a total value of $0.42(0.45)$ at 162 GPa and 0.39(0.41) at 250 GPa, when $N(\varepsilon)$($N(0)$) is used.

\begin{figure}
\includegraphics[width=8.4cm]{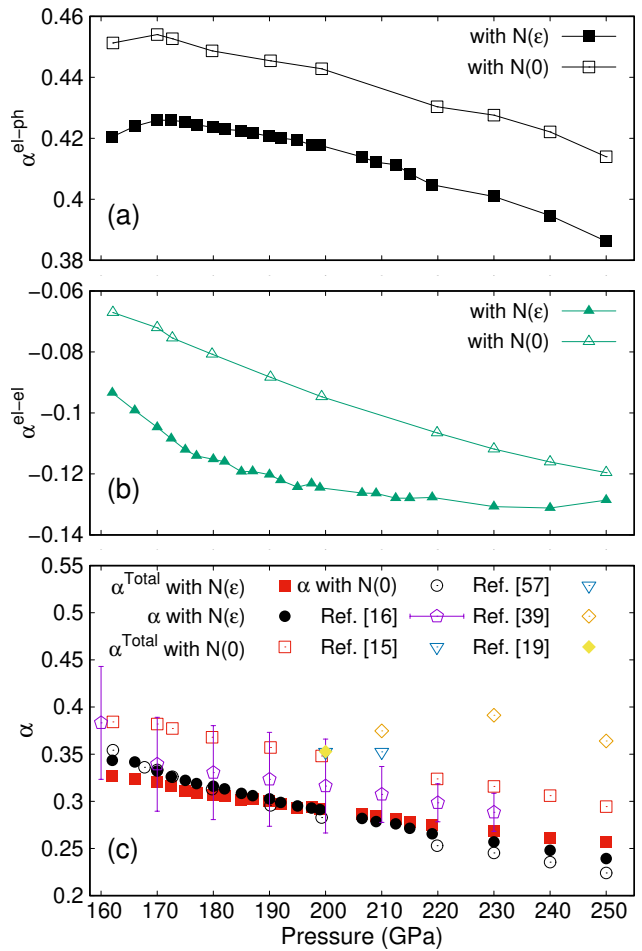}\caption{\label{fig:Iso_to}Calculated isotope effect coefficients within the Migdal-Eliashberg formalism. (a) Partial integrated electron-phonon $\alpha^{el-ph}$, (b) contribution from the electron-electron interaction, $\alpha^{el-el}$, and (c) the total isotope coefficient $\alpha$ calculated by Eq. (1) ($\alpha^{Total}$) and also by $\alpha=-\Delta\ln T_{c}/\Delta\ln M$, using our own calculated $T_{c}$. $\alpha$, obtained from the average of the experimental $T_{c}$ values reported in literature\cite{cor1}, is also presented, as well as theoretical results from previous publications\cite{Flores-Livas2016,PhysRevB.91.224513,CLm3m,PhysRevLett.114.157004}.
}
\end{figure}

In Fig. \ref{fig:Iso_to}(a) it is presented the partial integrated value $\alpha^{el-ph}$ for the entire range of studied pressure for both schemes, $N(\varepsilon)$ and $N(0)$. While the $N(0)$ case shows higher values than $N(\varepsilon)$, $\alpha^{el-ph}$ decreases steadily as a function of pressure, for both cases. Moreover, in Fig. \ref{fig:Iso_to}(b) it is shown the isotope effect contribution from the electron-electron interaction, $\alpha^{el-el}$. It indicates how the electron-electron screening is affected by the isotope substitution in H$_{3}$S. This parameter was calculated also for both LMEE versions ($N(\varepsilon)$ and $N(0)$). On both cases, $\alpha^{el-el}$ is negative with smaller absolute value than the electron-phonon contribution. While for 162 GPa its value is quite marginal, -0.09(-0.07) for $N(\varepsilon)$($N(0)$), it improves as pressure increases, reaching -0.13(-0.12) with $N(\varepsilon)$($N(0)$). In Fig. \ref{fig:Iso_to}(c) the total isotope coefficient $\alpha^{Total}$ as a function of pressure is presented, calculated by Eq. (\ref{eq:Totalalfa}) and also by $\alpha=-\Delta\ln T_{c}/\Delta\ln M$, using our own calculated $T_{c}$ values also on both schemes, with $N(\varepsilon)$ and $N(0)$. In general, both $\alpha$ descriptions follow the same reduction trend as a function of pressure. In particular, $\alpha$ calculated from $T_{c}$ shows small differences between $N(\varepsilon)$ and $N(0)$ schemes only for the high-pressure region (larger than 220 GPa). In comparison, $\alpha^{Total}$ with $N(0)$ overestimates the previous discussed $\alpha$ values by an amount as large as $24\%$ at 250 GPa. However, when $N(\varepsilon)$ is taken into account, a renormalization occurs, showing similar values as $\alpha$ coming from the calculated $T_{c}$, but with a less pronounced decrease as pressure arises. Finally, by comparing our results with experimental ones, obtained from the average of the experimental $T_{c}$ values reported in literature \cite{cor1}, it can be observed that although the $\alpha$ from $T_{c}$ agrees quite well (regardless of the scheme), at higher pressures they tend to apart from the experiment. Instead, for $\alpha^{Total}$ such agreement only improves as pressure arises, indicating the importance of taking into account the role of both, the electron-electron interaction, as well as the effects of the energy-range
variation on $N(\varepsilon)$.
\section{Conclusions}

Here we performed a detailed analysis of the superconducting isotope effect in H$_{3}$S in the framework of the Migdal-Eliashberg theory within two versions of the LMEE, with ($N(\varepsilon)$) and without ($N(0)$) energy-dependence on the density of states at the Fermi level. While the $N(0)$ version has been widely used to study the superconducting state in the high-$T_{c}$ metal hydrides, the $N(\varepsilon)$ approach includes explicitely the effects of pressure on van Hove singularities, an important feature that is present on the superconducting hydrides.

By density functional calculations, together with the functional derivative formalism, we determined the evolution of the critical temperature ($T_{c}$) of H$_{3}$S and D$_{3}$S on its $lm\bar{3}m$ crystal-structure range (from 162 to 250 GPa), getting, in general, a good agreement with $N(0)$, which is improved on the whole range of studied pressure when $N(\varepsilon)$ approach is applied. We found that the related changes from the electron-phonon interaction, $\Delta T_{c}^{el-ph}$, are the main responsible of the decrease of $T_{c}$ as a function of pressure. In addition, by taking into account $\Delta T_{c}^{N\left(\varepsilon\right)}$ the agreement with the experiments is improved, by compensating the strong reduction induced by the electron-phonon term.

Finally, to address the isotope effect coefficient calculation considered for $\alpha$ two main contributions: one positive coming from the electron-phonon (el-ph) and the other negative from the electron-electron interaction (el-el). $\alpha$ showed a monotonic reduction as a function of pressure, independent of applied scheme ($N(\varepsilon)$ or $N(0)$). However, when $N(\varepsilon)$ is taken into account, an important renormalization occurs on both contributions, el-ph and el-el, improving the agreement with experimental data, specially for the high-pressure regime. 

In view of these results, we conclude that taking into account the energy-dependence on $N(\varepsilon)$ its necessary for a proper analysis and description of the superconducting state on high-$T_{c}$ metal hydrides as H$_{3}$S, by considering the role of its van Hove singularities.

\begin{acknowledgments}
The authors thankfully acknowledge computer resources, technical advise, and support provided by Laboratorio Nacional de Superc{\'o}mputo del Sureste de M{\'e}xico (LNS), a member of the CONACYT national laboratories. One of the authors (S. Villa-Cort{\'e}s.) also acknowledges the Consejo Nacional de Ciencia y Tecnolog{\'i}a (CONACyT, M{\'e}xico) by the support under grant 741002.
\end{acknowledgments}

\appendix

\section{The functional derivatives\label{sec:The-functional-derivatives}}

In order to compute the functional derivatives, Eq. (\ref{eq:dev_a2f})
and Eq. (\ref{eq:dev_ne}), the LMEE have to be solved and the solution,
$\bar{\Delta}_{n}$, is used to get: 
\begin{equation}
\frac{\delta\rho}{\delta\alpha^{2}F\left(\omega\right)}=\frac{\sum_{nm}\bar{\Delta}_{n}\delta K_{nm}/\delta\alpha^{2}F\left(\omega\right)\bar{\Delta}_{m}}{\sum_{n}\bar{\Delta}_{n}^{2}},
\end{equation}
and
\begin{equation}
\frac{\delta\rho}{\delta N(\varepsilon)}=\frac{\sum_{nm}\bar{\Delta}_{n}\delta K_{nm}/\delta N(\varepsilon)\bar{\Delta}_{m}}{\sum_{n}\bar{\Delta}_{n}^{2}}.
\end{equation}
The equations needed to compute $\delta K_{nm}/\delta N(\varepsilon)$
were first given by Lie and Carbotte \cite{LIE1978511}:
\begin{equation}
\frac{\delta K_{nm}}{\delta N(\varepsilon)}=\frac{\partial K}{\partial\tilde{\omega}_{m}}\frac{\delta\tilde{\omega}_{m}}{\delta N(\varepsilon)}+\frac{\delta K}{\delta N(\varepsilon)},
\end{equation}
with 
\begin{equation}
\frac{\delta K}{\delta N(\varepsilon)}=\frac{T_{c}}{N\left(0\right)}\left[\lambda_{nm}-\mu^{*}\right]\frac{\left|\tilde{\omega}_{n}\right|}{\left|\tilde{\omega}_{n}\right|^{2}+\varepsilon^{2}},
\end{equation}
and 
\begin{equation}
\frac{\partial K}{\partial\tilde{\omega}_{m}}=\pi T_{c}\left[\lambda_{nm}-\mu^{*}\right]F_{m}-\delta_{nm}.
\end{equation}
For $\delta K_{nm}/\delta\alpha^{2}F\left(\omega\right)$, $K_{nm}$
has to be taken as an explicit function of $\alpha^{2}F\left(\omega\right)$
and $\tilde{\omega}_{m}$, which in turn depends on $\alpha^{2}F\left(\omega\right)$,
namely: 
\begin{equation}
K_{nm}\equiv K\left(\tilde{\omega}_{m},\alpha^{2}F\left(\omega\right)\right),
\end{equation}
and 
\begin{equation}
\tilde{\omega}_{n}\equiv Q_{n}\left(\tilde{\omega}_{m},\alpha^{2}F\left(\omega\right)\right),
\end{equation}
then
\begin{equation}
\frac{\delta K_{nm}}{\delta\alpha^{2}F\left(\omega\right)}=\frac{\partial K}{\partial\tilde{\omega}_{m}}\frac{\delta\tilde{\omega}_{m}}{\delta\alpha^{2}F\left(\omega\right)}+\frac{\delta K}{\delta\alpha^{2}F\left(\omega\right)},
\end{equation}
with 
\begin{equation}
\frac{\delta K}{\delta\alpha^{2}F\left(\omega\right)}=\pi T_{c}\tilde{N}(\left|\tilde{\omega}_{m}\right|)\frac{\delta\lambda_{nm}}{\delta\alpha^{2}F\left(\omega\right)},
\end{equation}
and 
\begin{equation}
\frac{\partial K}{\partial\tilde{\omega}_{m}}=\pi T_{c}\left[\lambda_{nm}-\mu^{*}\right]F_{m}-\delta_{nm}.
\end{equation}
Here, the functional derivative of $\lambda_{nm}$ is given by 
\begin{equation}
\frac{\delta\lambda_{nm}}{\delta\alpha^{2}F\left(\omega\right)}=\frac{2\omega}{\omega^{2}+\left(\omega_{n}-\omega_{m}\right)^{2}},
\end{equation}
and $F_{m}$ is defined as 
\begin{equation}
F_{m}\equiv\frac{\partial\tilde{N}(\left|\tilde{\omega}_{m}\right|)}{\partial\left|\tilde{\omega}_{m}\right|}=\int_{-\infty}^{\infty}d\varepsilon\frac{N\left(\varepsilon\right)}{N\left(0\right)}\frac{\varepsilon^{2}-\left|\tilde{\omega}_{m}\right|^{2}}{\left(\varepsilon^{2}+\left|\tilde{\omega}_{m}\right|^{2}\right)^{2}}.
\end{equation}
To compute $\delta\tilde{\omega}_{m}/\delta\alpha^{2}F\left(\omega\right)$
we have to solve the following set of equations 
\begin{equation}
\frac{\delta Q_{n}}{\delta\alpha^{2}F\left(\omega\right)}=\sum_{m}\left[\delta_{nm}-\frac{\partial Q_{n}}{\tilde{\omega}_{m}}\right]\frac{\delta\tilde{\omega}_{m}}{\delta\alpha^{2}F\left(\omega\right)},
\end{equation}
where 
\begin{equation}
\frac{\partial Q_{n}}{\tilde{\omega}_{m}}=\pi T_{c}\lambda_{nm}\textrm{Sig}\left(\omega_{m}\right)F_{m},
\end{equation}
and 
\begin{equation}
\frac{\delta Q_{n}}{\delta\alpha^{2}F\left(\omega\right)}=\pi T_{c}\sum_{m}\frac{\delta\lambda_{nm}}{\delta\alpha^{2}F\left(\omega\right)}\textrm{Sig}\left(\omega_{m}\right)\tilde{N}(\left|\tilde{\omega}_{m}\right|).
\end{equation}
Finally, the derivative of $\rho$ with respect to $T$ at the critical
temperature 
\[
\left.\left(\frac{d\rho}{dT}\right)\right|_{T_{c}},
\]
in Eqs. (\ref{eq:dev_a2f}) and (\ref{eq:dev_ne}) is calculated numerically
solving the LMEE. 

\bibliographystyle{apsrev4-1}
\bibliography{H3S_ISO}

\end{document}